\documentclass[aps,prl,twocolumn,nofootinbib,preprintnumbers]{revtex4-1}
\usepackage{bm}
\usepackage{hyperref}
\usepackage{amsmath,amssymb,amsfonts,graphicx,float}

\newcommand{\beq}{\begin{eqnarray}}
\newcommand{\eeq}{\end{eqnarray}}
\usepackage{amsmath}
\usepackage[paper=letterpaper,margin=1in]{geometry}
\usepackage{braket}
\usepackage{upgreek}

\def\G{\Gamma}

\usepackage{tikz}
\usepackage{subfigure}

\begin{document}

\title{Divergent Thermopower without a Quantum Phase Transition}
\author{Kridsanaphong Limtragool and Philip W. Phillips}

\affiliation{Department of Physics and Institute for Condensed Matter Theory,
University of Illinois
1110 W. Green Street, Urbana, IL 61801, U.S.A.}

\date{\today}

\begin{abstract}
A general principle of modern statistical physics is that divergences
of either thermodynamic or transport properties are only possible if
the correlation length diverges.  We show by explicit calculation that  the thermopower in the quantum XY model $d=1+1$ and the Kitaev model in $d=2+1$ 
can
1) diverge even when the correlation length is finite and 2) remain finite
even when the correlation length diverges, thereby providing a counterexample to the 
standard paradigm.Two conditions are necessary: 1) the sign of the charge carriers and that
of the group velocity must be uncorrelated and 2) the 
current operator defined formally as the derivative of the Hamiltonian
with respect to the gauge field does not describe a set of excitations
that have a particle interpretation, as in strongly correlated
electron matter. The recent
experimental\cite{2dtp} and theoretical\cite{kirkpatrick} findings on
the divergent thermopower of a 2D electron gas are discussed in this context.

\end{abstract}
\maketitle

A truism in modern statistical mechanics is that divergences (or more
generally non-analyticities)
in a thermodynamic quantity or a transport property always signal a transition
to a new state of matter.  In fact, the very notion of adiabatic continuity is based on
the intuition that non-analyticities resulting from tuning some system
parameter cannot emerge without the crossing of a phase boundary.
More precisely, as long as the correlation length remains finite, then
no divergences are possible because both transport and thermodynamic
properties are governed by the singular part of the free energy.  We
present here a counter example to this rule. To establish our result, we consider the quantum XY model in 1D and the Kitaev\cite{Kitaev20062}
model in 2D, both of which can be solved\cite{katsura,jordanwignerk,ortiz} exactly using a mapping to
fictitious fermionic degrees of freedom.  In both cases, we show exactly that the 
thermopower, appropriately defined, diverges  at
fillings that have nothing to do with the quantum phase transition in these models. At the
spurious divergences, no thermodynamic quantity experiences a
non-analyticity. As we will see, the heart of this problem is a breakdown of the particle
interpretation of the current-carrying degrees of freedom.

This work is motivated by recent measurements\cite{2dtp} on the thermopower
in a dilute 2D electron gas. These experiments are the latest in a
series of remarkable observations\cite{2dmit} that a dilute 2D electron gas
exhibits a resistivity that decreases as the temperature is lowered with no
apparent upturn (as is expected from the scaling theory\cite{go4}) thereby providing evidence for a low-temperature
metallic state.  Mokashi et al. reported\cite{2dtp} that the thermopower 
on the metallic side of the transition diverges
exhibiting scaling of the form
\beq\label{thermoscal}
&& S(T,n)=eT s(n)=T(n-n_c)^{-\mu}
\eeq
with $\mu=1.0\pm 0.1$. 
Consequently, if the thermpower were to be measured on the insulating
side, it should change sign.
As a result, they
interpreted\cite{2dtp} such a critical divergence, based on a simple appeal to
the adiabatic continuity principle, as definitive evidence that the
transition to the metallic state represents a true $T=0$ quantum phase
transition.  This would then represent the most important
finding since the initial discovery paper in 1996\cite{1996}.  More recently, Kirkpatrick and Belitz\cite{kirkpatrick}
argued that the divergence of the thermopower holds crucial
implications for the scaling of the specific heat as the exponent $\mu$
determines the product of dynamical and correlation length exponents,
$z$ and $\nu$, respectively. 

Hence, while explaining the experimental data
is certainly of interest, our focus is on whether alternative
mechanisms exist for a divergent thermopower other than a quantum phase transition.
Although the models in the counterexamples we construct  are not directly applicable to the experiments,
the mechanism for the divergence of the thermopower is.  We find that in strongly correlated systems, the thermopwer can diverge simply because the 
the current does not have a particle interpretation.

We treat at first the quantum
XY model in 1D.  This model can be fermionized\cite{katsura,sachdev} 
\beq\label{eq1}
H &=& -\sum\limits_{i}(c_i^\dagger c_{i+1}+c_{i+1}^\dagger c_i + \Gamma c_i^\dagger c_{i+1}^\dagger + \Gamma c_{i+1}c_i \nonumber\\&+& h(1-2c_i^\dagger c_i)) \label{eq:xy_H},
\eeq
using the Jordan-Wigner
transformation scheme with $c_i$ a canonical fermionic annihilation
operator for site $i$.  The hopping between two sites is set to 1 in the unit of
$J = \frac{1}{2}(J_x+J_y)$, $\Gamma =
\frac{J_x-J_y}{2J}$ is a measure of the exchange anisotropy and  $h =
\frac{H}{J}$ is the effective magnetic field or in the fermionic model
$-2h$ is a dimensionless chemical potential.  Although $\Gamma\ne0$
implies an effective particle non-conservation, thereby making it
possible to fix only the average number of particles, we have shown\cite{dave} that 
a unique expression\footnote{ Although 4 expressions (Eqs. (6a-6d)) are derived in Ref. [11] for the thermopower,
Eqs. (6c) and (6d) are valid
only for a transverse field, while Eqs. (6a) and (6b) apply strictly for a
longitudinal field.  (6a) follows from (6b) from the continuity equation which is not valid here.  Since the thermopower experimentally is the response to a longitudinal field, only Eq. (6b) is valid in this context and hence the thermopower has a unique definition.}  exists for the thermopower defined as a response to a longitudinal field.   We calculated the exact
expression for the thermopower\cite{dave} and showed that it diverges
at the phase transition,  $h=\pm1$.  However, this analysis is far from complete as
we will show below.  There are additional divergences away from $h=\pm1$
at which the thermodynamics is completely smooth. 

To analyze the thermopower, we Fourier transform the Hamiltonian and
diagonalize it using a Bogoliubov transformation.  The diagonalized
Hamiltonian,
\beq
H = \sum\limits_{k}\varepsilon_k\gamma_k^\dagger \gamma_k \label{eq:H_diagonal},
\eeq
contains the new fermionic operators, $
\gamma_k = u_kc_k - iv_kc_{-k}^\dagger$ and
$\gamma_k^\dagger = u_kc_k^\dagger + iv_kc_{-k}$,
whose energies are $\varepsilon_k = \pm2\sqrt{(h-\cos
  k)^2+\Gamma^2\sin^2k}$
with $u_k = 2\cos\frac{\theta_k}{2}$ and $v_k = 2\sin\frac{\theta_k}{2}$
and the angle $\theta_k$ defined through
$\sin \theta_k = (\Gamma \sin k)/\varepsilon_k$ and $\cos \theta_k =(h-\cos k)/\varepsilon_k$.
We will be analyzing the properties of this model as a function of the
average particle density,
\beq
x= \langle c_i^\dagger c_i \rangle =\frac{1}{2\pi}\int_0^\pi dk\left(1-\cos\theta_k\tanh\left(
\frac{\beta|\varepsilon_k|}{2}\right)\right).\nonumber
\eeq

The thermodynamic quantity of interest is the heat capacity,
\beq
\frac{C}{N} =\frac{k_B}{4\pi}\int\limits_{0}^{\pi}(\frac{\varepsilon_k}{k_BT})^2 \mathrm{sech}^2(\frac{\beta\varepsilon_k}{2}).
\eeq

\begin{figure}
\centering
\includegraphics[scale=0.7]{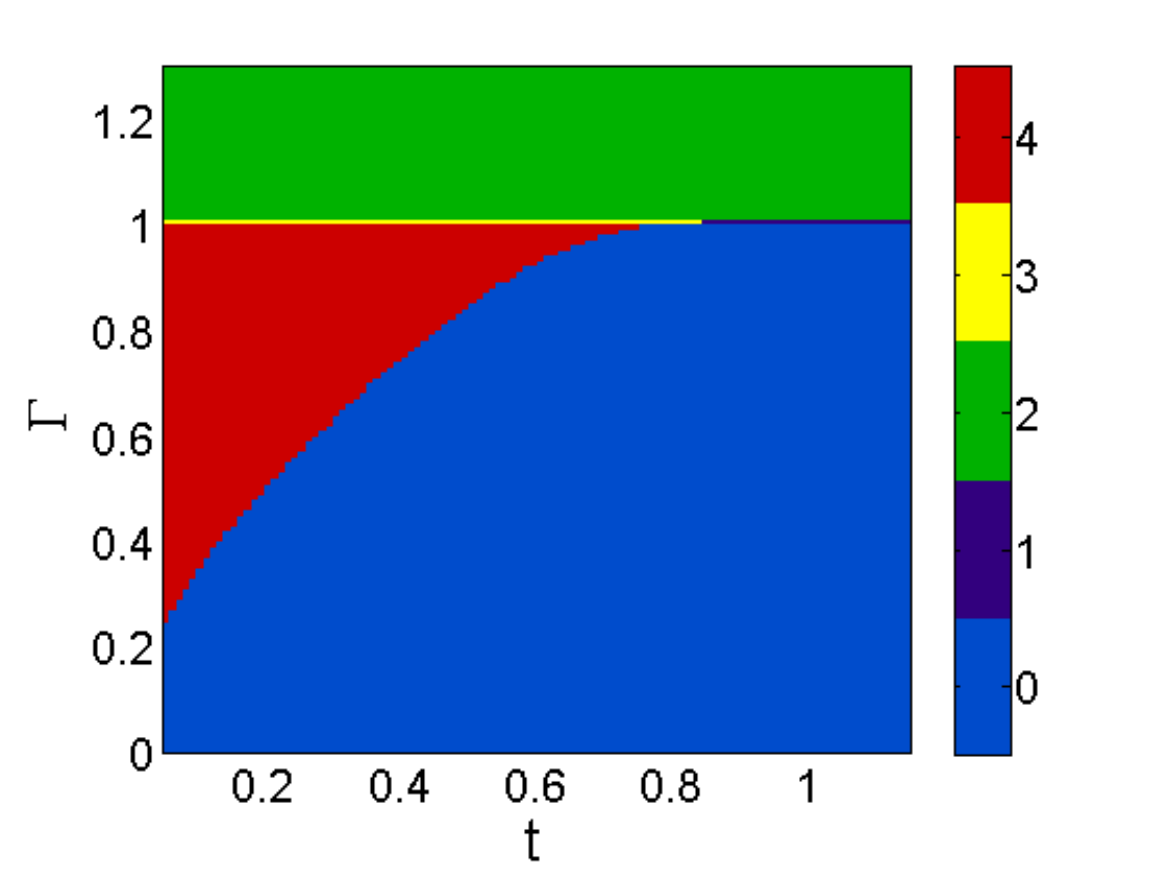}
\caption{This is a plot of a number of divergences in thermopower vs. particle filling at given values of $\Gamma$ and $t$ in the fermionized quantum XY model.}
\label{fig:1d phase}
\end{figure}

However, our main focus is the thermopower. To this end, we write the
charge ( $\hat{J}_x$) and thermal currents ($\hat{J}_x^Q$) along the
x-direction in terms\cite{shastry} of the responses to an electric field and a
temperature gradient,
\beq
\frac{1}{\Omega}\langle \hat{J}_x \rangle &=& L_{11}E_x + L_{12}\bigg(-\frac{\nabla_xT}{T}\bigg) \label{eq:current} \\
\frac{1}{\Omega}\langle \hat{J}_x^Q \rangle &=& L_{21}E_x + L_{22}\bigg(-\frac{\nabla_xT}{T}\bigg) \label{eq:heat_current}
\eeq
 using the Onsager coefficients, $L_{ij}$. In these expressions,
 $\Omega$ is the volume of the system.  The thermopower\cite{shastry},
\beq
Q = \frac{L_{12}}{TL_{11}}, \label{eq:Q2}
\eeq
is the ratio of the voltage generated per gradient of temperature.
An explicit calculation of $L_{ij}$ is possible in frequency and
momentum space for the models we consider here.  The transport or fast limit corresponds to 
$\lim_{\omega\rightarrow 0}\lim_{q_x\rightarrow 0}$. For the quantum
XY model in 1D, the exact expression\cite{dave} for the thermopower,
\beq
\lim_{\omega\rightarrow 0}\lim_{q_x\rightarrow 0}\frac{eQ}{k_B} = \frac{\int\limits_{-\pi}^{\pi}dk\frac{\varepsilon_k}{k_BT} \sin k \frac{dn}{dk}}{\int\limits_{-\pi}^{\pi}dk(u_k^2-v_k^2)\sin k\frac{dn}{dk}} \label{eq:thermopower_xy},
\eeq
involves a simple integral over the first Brillouin zone with an
integrand determined by the coherence factors and the fermionic
occupation, $n=1/(e^{\varepsilon_k/k_BT}+1)$.
\begin{figure}
      \centering
      \subfigure[$XY$ model with $\Gamma = 0.8, t = 0.1$ \label{xy_g08_t01}]{\includegraphics[scale=0.6]{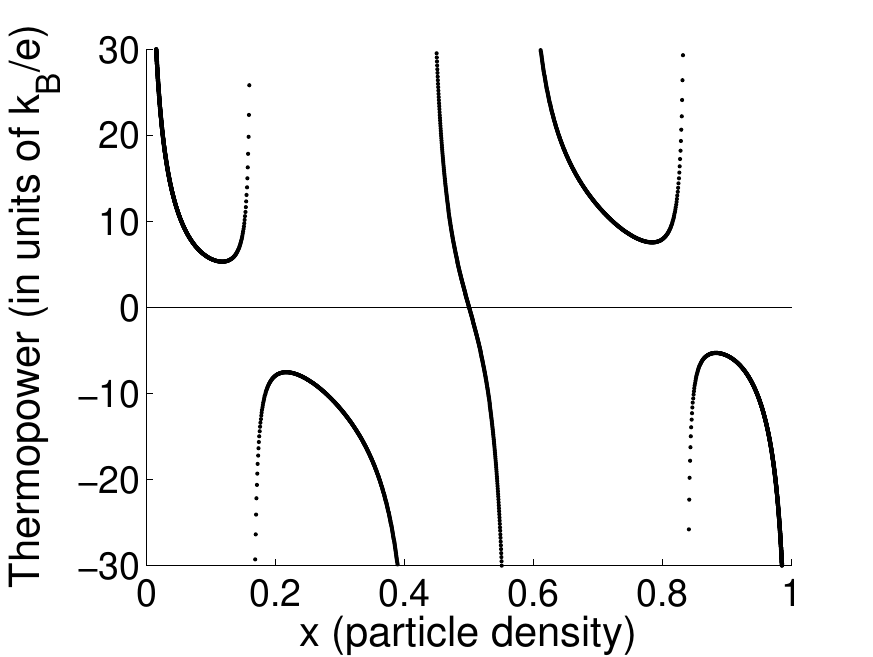}}
        \subfigure[Kitaev model with $J_y = 0.2, t = 0.3$ \label{kitaev_jy02_t03}]{\includegraphics[scale=0.6]{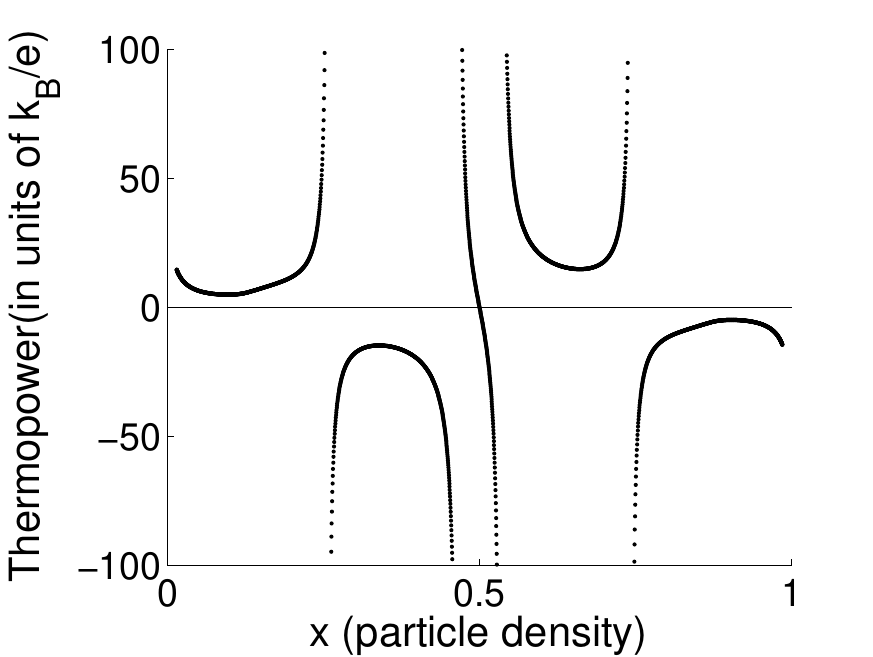}}
       \caption{The plots of the thermopower vs. particle density. }
       \label{thermo1}
\end{figure}
The numerator of this expression is bounded over integration in the first
Brillouin zone.  Consequently any divergence arises entirely from the
denominator.   We display the results for $\G=0.8$ and $t=0.1$ in the first panel in Fig. (\ref{xy_g08_t01}), where $t$ is the dimensionless temperature and defined as $t=k_B T/J$.  For these parameters, the particle density
at the phase transition, $h=\pm1$, is $x\approx 0.15$ or the
particle-hole complement, $x\approx .85$. Fig. \ref{xy_g08_t01} shows that indeed
the thermopower does diverge at these values of $x$ as we reported
earlier\cite{dave}.  However, there are other divergences, for example at $x\approx0.4,0.6$ in Panel \ref{xy_g08_t01}, away from the critical value of the
filling. The full phase diagram for this
model in terms of the total number of divergences is catalogued in
Fig. (\ref{fig:1d phase}).  There are a total of five regions: a) no
divergences (blue), b) four divergences (red, Panel \ref{xy_g08_t01}), c) three
divergences (yellow), d) two divergences (green), and
e) one divergence (purple).    Fig. (\ref{thermodynamics})  illustrates that only at the
phase transition does the heat capacity display the characteristic peak-like feature which turns into a non-analyticity at $T=0$.
Hence, non-analyticities in thermodynamics need not
affect transport properties and conversely divergences in transport
properties are not necessarily accompanied by singularities in
the thermodynamics.
Before we analyze the origin of these results, we first show that our
findings are not an artifact of 1-dimensional (d=1+1) physics.  To this
end, we consider the Kitaev model,
\beq
H &=& -J_x\sum\limits_{\mathrm{x-bonds}}\sigma_R^x\sigma_{R'}^x -
J_y\sum\limits_{\mathrm{y-bonds}}\sigma_R^y\sigma_{R'}^y\nonumber\\
& - &J_z\sum\limits_{\mathrm{x-bonds}}\sigma_R^z\sigma_{R'}^z,
\eeq
on a honeycomb lattice in which the summations are over all links between site $R$ and $R'$. This Hamiltonian can be fermionized\cite{jordanwignerk} by the Jordan-Wigner transformation. The result is a model of Dirac fermions, 
\beq
H &=& J_x\sum\limits_{i}(c_i^\dagger + c_i)(c_{i+\hat{x}}^\dagger -
c_{i+\hat{x}})+ J_y\sum\limits_{i}(c_i^\dagger + c_i)\nonumber\\
&\times&(c_{i+\hat{y}}^\dagger - c_{i+\hat{y}}) + J_z\sum\limits_{i}\alpha_i(2c_i^\dagger c_i - 1),
\eeq
on a square lattice.
At every lattice site there is one conserved quantity, $\alpha_i$,
which has the value of -1 or 1. The ground state of this system corresponds to having $\alpha_i$
equal to 1 everywhere. So we choose all $\alpha_i$ to be
1. This Hamiltonian can be solved exactly in the
same way as the quantum XY model\cite{jordanwignerk,ortiz}.
The energy spectrum is given by $\varepsilon_k =
2\sqrt{(J_z-\sum_iJ_i\cos k_i)^2 + (\sum_i J_i \sin k_i)^2}$ and the
coherence factors defined through
 the parameters $u_k$ and $v_k$ in the Bogoliubov transformation satisfy
\beq
 \cos\theta_k &=& u_k^2 - v_k^2 = \frac{2(J_z-J_x\cos k_x - J_z \cos k_y)}{\varepsilon_k} \nonumber \\
 \sin\theta_k &=& 2u_kv_k = \frac{2(J_x \sin k_x + J_y \sin
   k_y)}{\varepsilon_k} .
\eeq
The sum on $i$ in the energy spectrum above is over $x$ and
$y$. The analogous expression for the thermopower,
\beq
\frac{eQ}{k_B} = \frac{\int\limits_{-\pi}^{\pi}dk_x \int\limits_{-\pi}^{\pi}dk_y \frac{\varepsilon_k}{k_BT} \sin k_x \frac{dn}{dk_x}}{\int\limits_{-\pi}^{\pi}dk_x \int\limits_{-\pi}^{\pi}dk_y (u_k^2-v_k^2)\sin k_x\frac{dn}{dk_x}},
\label{eq:thermopower_kitaev}
\eeq
obtained from an exact calculation of $L_{ij}$ in the fast limit, is
precisely the 2D generalization of Eq. (\ref{eq:thermopower_xy}).
\begin{figure}
\centering
\includegraphics[scale=0.7]{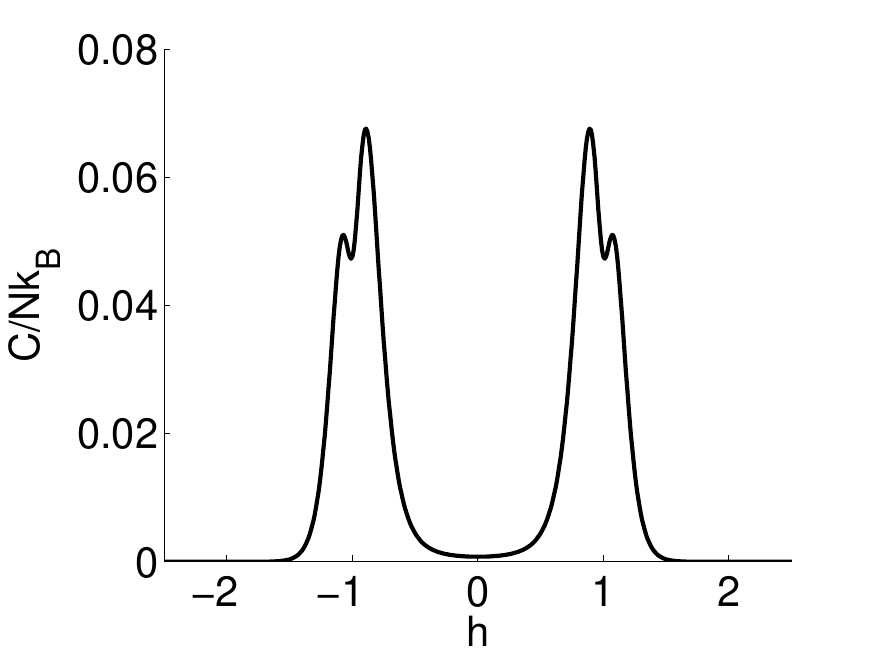}
\caption{Heat capacity in the quantum XY model at parameter values $\G=0.8$ and $t=0.1$ clearly shows a non-analyticity at the quantum phase
  transition $h=\pm1$ or $x\approx0.15,0.85$.  At $h=\pm0.27$ or $x\approx0.4,0.6$, thermopower diverges but there is no thermodynamic signature at these points (see Fig. \ref{xy_g08_t01}).}
\label{thermodynamics}
\end{figure}

\begin{figure}
\centering
\includegraphics[scale=0.7]{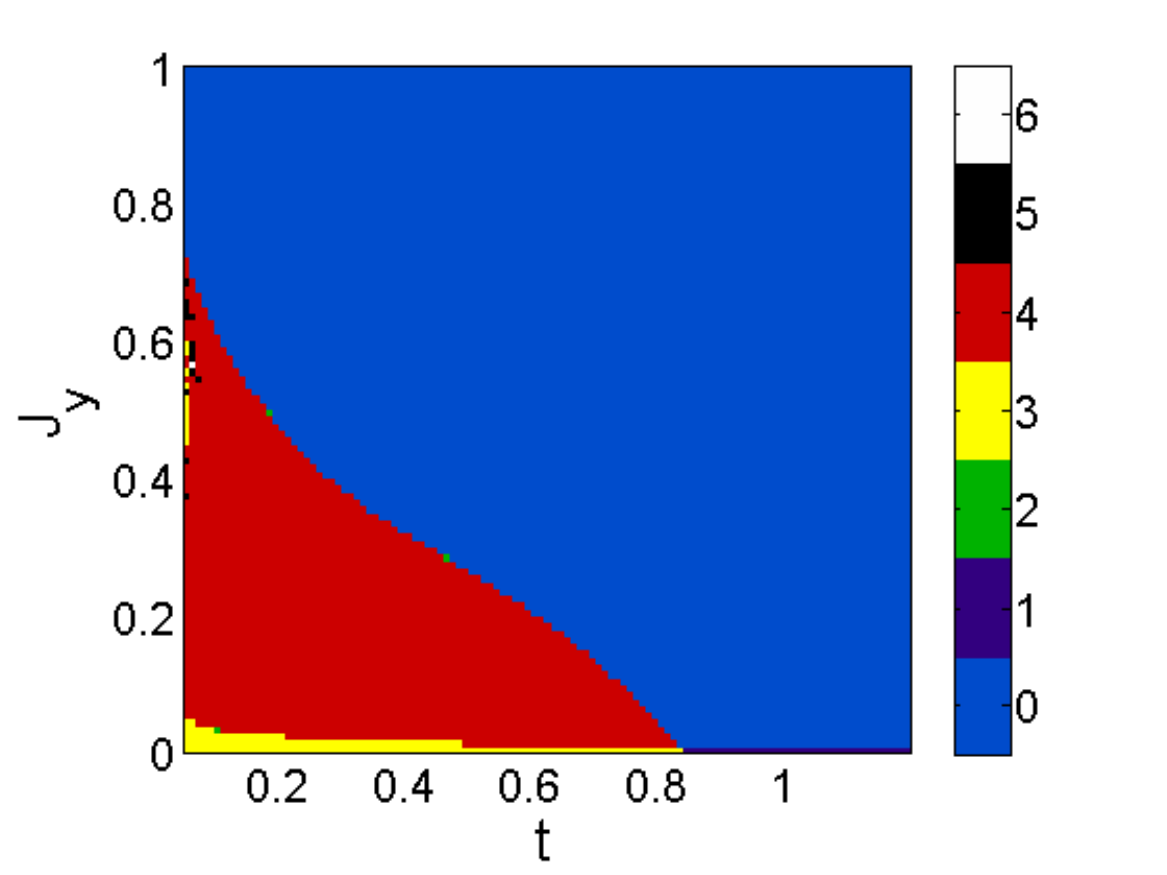}
\caption{Number of divergences in the thermopower versus
  particle filling for fixed values of $J_y$ and $t$ of the
  fermionized Kitaev model. Each color region displays a different
  number of divergences. }
\label{fig:kitaev phase}
\end{figure}

For the Kitaev model, the thermopower, $Q = Q(J_x,J_y,J_z,t)$, depends
on the average particle density $x = x(J_x,J_y,J_y,t)$. We write $J_y$
and $J_z$ in units of $J_x$ (by setting $J_x = 1$). So for a fixed
value of $J_y$ and $t$, we can plot thermopower versus particle
density by varying $J_z$. Figs. (\ref{kitaev_jy02_t03}) and (\ref{fig:kitaev phase}) demonstrate
that the behaviour is identical to that of the quantum XY
model. Hence, our results are not an artifact of 1-dimensional
physics.   Note that this model also exhibits regions in which no
divergence obtains although the quantum phase transition is present.  

\begin{figure}
\centering
		\subfigure[$\Gamma=0.8,t=0.1,h=0.27$]{\includegraphics[scale=0.60]{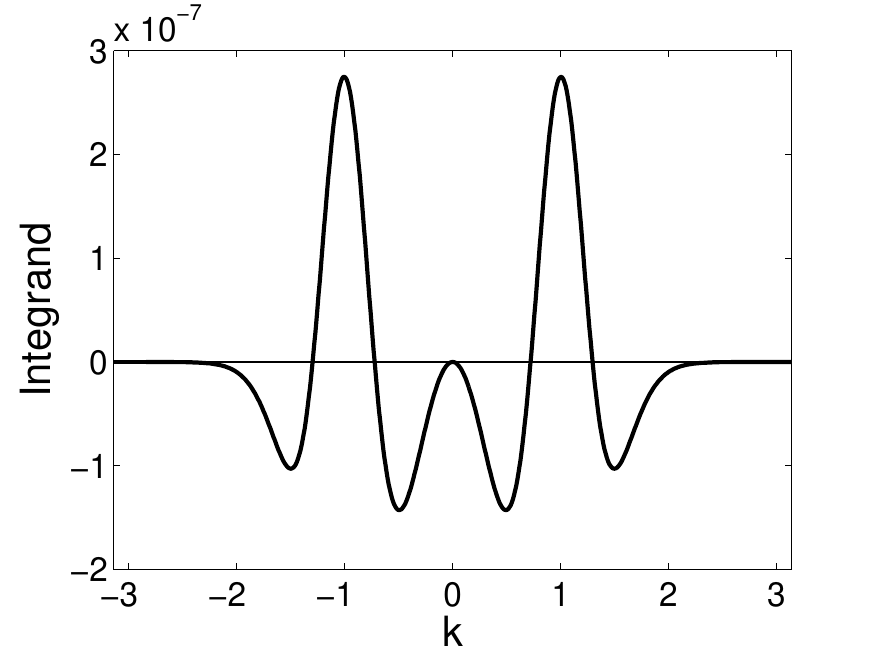}}
\caption{Integrand of $L_{11}$ (denominator of the thermopower) showing the cancellation which leads to a divergence in the thermopower. }
\label{cancellation}
\end{figure}

The origin of this physics is tied to the denominators of the
expressions for the thermopower because $L_{12}$ is a completely
bounded function  for all values of $k$ inside the first Brillouin
zone. Consider the denominator, in the case of the XY model
\beq\label{L11int}
\mathrm{XY} &\rightarrow& \int\limits_{-\pi}^{\pi}dk(u_k^2-v_k^2)\sin
k\frac{dn}{dk},
\eeq
the Kitaev
model being the direct 2D analogue. 
The $\sin k$ factor arises from the momentum dependence of the local current
operator, $J_j=-i(c_j^\dagger c_{j+1}-c^\dagger_{j+1}c_j)$. The
quantity $q_k=u_k^2-v_k^2=\cos\theta_k\propto h-\cos k$ is the
effective charge of the
quasiparticles, which is even with respect to $k$.  It is instructive
then to rewrite the denominator, 
\beq
I=\int_{-\pi}^{\pi} dk J(k) n_{k+v_d},
\eeq
in a form which lays plain that it is no more than the current in
response to the applied field with $v_d=q_k E_x \tau$, the drift velocity, and
$J(k)$ the momentum dependence of the current operator.  In the absence of the drift velocity, $I=0$.
Taylor expanding around $v_d=0$ yields Eq. (\ref{L11int}). Herein lies
the crux of the problem.  In a non-interacting system, the
local definition of the current operator used here and that arising
from the continuity equation both yield the same result, namely that $J(k)=q_k d\varepsilon_k/dk=dH/dk$, in which
case the integrand is positive definite and cannot integrate to
zero. However, for the problem at hand, the current operator arising from the
continuity equation, namely $q_kd\varepsilon_k/dk$, is non-local in space, possessing sink and source terms, and hence
is not tenable.  Such non-locality typifies
 most strongly correlated systems because the entities which carry
the current are not simply determined by the kinetic part of the
Hamiltonian.  Consequently, the current operator, defined from the
continuity equation is non-local and lacks a particle interpretation.
In such cases, $I$ can vanish.  The vanishing of $I$ here takes place
because the group
velocity, $d\varepsilon_k/dk$,
is an odd function of $k$, while $q_k$ is even.  Consequently, the momenta at which $q_k$
and $d\varepsilon_k/dk$ change sign need not be correlated. Because the overall
integrand is an even function of $k$, it will have positive and
negative contributions on the interval $[0,\pi]$, which for certain
system parameters could yield a cancellation as illustrated in
Fig. (\ref{cancellation}).  

Classic examples in which the operators in the local current operator do not coincide
with the charge carriers are the insulating state of the Hubbard model
at half-filling for sufficiently large $U$.  In this problem, there is
no divergent length scale as there is no order parameter for the Mott
insulating state.  It is entirely likely that the insulator in the dilute 2D
electron gas\cite{2dtp} is induced by the correlations as well as it obtains in
the large $r_s$ regime.  Hence, caution must be taken in
using standard scaling arguments to relate the thermopower to
divergent correlation lengths as has been done recently\cite{kirkpatrick}.  Unless the
charge carriers are local degrees of freedom, naive scaling with the
correlation length is insufficient to describe transport properties
such as the thermopower.

\textbf{Acknowlegements} We thank Taylor Hughes, Jeffrey Teo, Mike Stone, Brandon Langley, Tony Hegg, Wei-cheng Lee, Ted Kirkpatrick, and Nigel
Goldenfeld for sustained commentary throughout the completion of this
work and NSF DMR-1104909 for partial funding of this project. KL is supported by the Department of Physics at the University of Illinois and a scholarship from the Ministry of Science and Technology, Royal Thai Government.

\end{document}